# A BAYESIAN ASSESSMENT OF P-VALUES FOR SIGNIFICANCE ESTIMATION OF POWER SPECTRA AND AN ALTERNATIVE PROCEDURE, WITH APPLICATION TO SOLAR NEUTRINO DATA

P.A. STURROCK

Center for Space Science and Astrophysics, Stanford University, Stanford, CA 94305

J. D. SCARGLE

NASA/Ames Research Center, MS 245-3, Moffett Field, CA 94035

ABSTRACT

The usual procedure for estimating the significance of a peak in a power spectrum is to calculate the probability of obtaining that value or a larger value by chance (known as the "p-value"), on the assumption that the time series contains only noise - typically that the measurements are derived from random samplings of a Gaussian distribution. However, since the use of p-values in other contexts is known to be misleading, it seems prudent to examine the implications of using p-values for significance estimation of power spectra.

We really need to know the probability that the time series is – or is not – compatible with the "null hypothesis" that the measurements are derived from noise. This probability can be calculated by Bayesian analysis, but this requires one to specify and evaluate a second hypothesis, that the time series does contain a contribution other than noise. We show that the requirement that the p-value should be identical to the probability that the null hypothesis is true leads to an unacceptable form for the likelihood function associated with this hypothesis. We claim that, for this reason, the p-value is not an acceptable method for significance estimation of a power-spectrum.

In order to obtain an acceptable significance estimate, it is necessary to consider explicitly a second hypothesis, and the key challenge is to identify an appropriate likelihood function for this hypothesis. We approach the problem of identifying this function in two ways. We first propose three simple conditions that it seems reasonable to impose on this function, and show that these conditions may be satisfied by a simple function with one free parameter.

We then define two different ways of combining information derived from two independent power estimates. One procedure is to calculate the post-probabilities of the null hypothesis, convert these to odds-values, and sum the odds. The second procedure is to combine the p-values using a procedure due to R.A. Fisher, and to calculate the corresponding post-probability and hence the corresponding odds. It seems sensible – even if not logically essential - to seek a likelihood function for which the two procedures lead to the same answer. We find that this consistency condition may be satisfied, to good approximation, by a special case of the previously proposed likelihood function.

We find that the resulting significance estimates are considerably more conservative than those usually associated with the p-values. As two examples, we apply the new procedure to two recent analyses of solar neutrino data: (a) power spectrum analysis of Super-Kamiokande data, and (b) the combined analysis of radiochemical neutrino data and irradiance data.

1. INTRODUCTION

The usual approach to significance estimation of power spectra is to compute the probability of obtaining the specified power or more on the basis of the null hypothesis that the time series consists only of noise. For the familiar assumption that the noise is Gaussian, the likelihood of obtaining power S in the range $S\ to\ S + dS$ is given by

$$P_S(S|H0)dS = e^{-S}dS\ ,\qquad(1)$$

and the likelihood of getting S or more is given by

$$P_>(S|H0) = \int_S^\infty dz\, P_S(z|H0) = e^{-S}\ .\qquad(2)$$

(See, for instance, Groth 1975, Scargle 1982.) Conventional significance estimates are based on Equation (2).

The probability of getting a certain result or a "more extreme" result on the basis of a "null hypothesis" (as in Equation (2)) is known as the "p-value." Textbooks on statistics (see, for instance, Utts, 1996) emphasize that **a p-value should not be interpreted as the probability that the null hypothesis is false.** See also Goodman (1999). We have previously investigated the use of p-values in relation to the Bernoulli (coin flipping) problem (Sturrock 1997) and found that it is misleading as a significance estimator.

We begin, in Section 2, by adopting a Bayesian approach (see, for instance, Good, 1983a,b; Howson and Urbach, 1989; Jaynes, 2004; Sturrock, 1973, 1994), the key point of which is that we must specify a complete set of hypotheses, not just a single hypothesis. It is important to note that **this analysis begins with an estimate of the power derived from power-spectrum analysis of a time series, not with the time series itself.** As in our analysis of procedures for combining power estimates (Sturrock, Scargle, Walther, and Wheatland, 2005), we here assume that we are given the results of a power-spectrum analysis, but we have no access to the time series from which those results were derived.

In Section 3, we demonstrate that a p-value may not be interpreted as an estimate of the probability, based on the power spectrum, that the null hypothesis (that the time series is derived exclusively from Gaussian noise) is correct. For this reason, we proceed tentatively to look for an alternative procedure for assessing the significance of peaks in a power spectrum.

We denote by $P_S(S|H1)dS$ the likelihood of finding S to be in the range $S$ to $S+dS$ if a signal is present. Our challenge is to determine an appropriate form for $P_S(S|H1)$. In Section 4, we propose what we consider to be four reasonable requirements for a functional form for the likelihood, and we propose a formula, with one free parameter, that meets these requirements. In Section 5, we examine a further constraint on the form of the likelihood function: we require that two different ways of combining power estimates (a Bayesian procedure and a Fisherian procedure) should lead to the same result. This may not be a logically essential requirement, but it is certainly a plus rather than a minus. This further requirement enables us to fix what had been a free parameter. We discuss this result and examine its implications for recent analyses of solar neutrino data in Section 6.

It should be emphasized that the main purpose of this article is to argue against the use of p-values for significance estimation of power series, and so to stress the need for something better. We do not pretend to have the final answer as to what that "something better" may be. We make a specific proposal primarily with the goal of "jump-starting" the search (to use a crude but familiar metaphor).

## 2. BASIC EQUATIONS

If we can determine the probability distribution function $P_S(S|H1)$ for S, we may derive by Bayes' theorem (Good, 1983a,b; Howson and Urbach, 1989; Jaynes, 2004; Sturrock, 1973, 1994) the probability that the time series contains a periodic signal

$$P(H0|S) = \frac{P_S(S|H0)}{P_S(S|H0)P(H0|-) + P_S(S|H1)P(H1|-)} P(H0|-), \qquad (3)$$

where $P(H0|-)$ and $P(H1|-)$ are the prior probabilities of H0 and H1.

In the absence of information that would lead one to favor H0 over H1 or vice versa, it is convenient to assign equal prior probabilities to H0 and H1, so that Equation (3) becomes

$$P(H0|S) = \frac{P_S(S|H0)}{P_S(S|H0) + P_S(S|H1)}. \qquad (4)$$

Similarly,

$$P(H1|S) = \frac{P_S(S|H1)}{P_S(S|H0) + P_S(S|H1)}. \qquad (5)$$

(Note that we now need consider only the actual value of the power; it is not necessary to consider the probability that the power is "S or more.")

Introducing the symbol $\Omega$ to denote "odds", we see that since H1 is the same as "not-H0," the "odds" on H0, based on the measurement S, is given by

$$\Omega(H0|S) \equiv \frac{P(H0|S)}{P(H1|S)} = \frac{P_S(S|H0)}{P_S(S|H1)}. \qquad (6)$$

In terms of "log-odds" (Good, 1983a,b), defined by

$$\Lambda(H0\,|\,S) = \log_{10}(\Omega(H0\,|\,S)), \tag{7}$$

Equation (6) becomes

$$\Lambda(H0\,|\,S) = \log_{10}\left(\frac{P_S(S\,|\,H0)}{P_S(S\,|\,H1)}\right). \tag{8}$$

The odds for H1 is the inverse of the odds for H0, and the log-odds for H1 is the negative of the log-odds for H0. We may if necessary retrieve the probability from the odds as follows:

$$P = \frac{\Omega}{\Omega+1}. \tag{9}$$

## 3. THE INADEQUACY OF P-VALUES FOR POWER-SPECTRUM SIGNIFICANCE ESTIMATION

We now wish to investigate whether there is any form of $P_S(S\,|\,H1)$, the likelihood of S on the basis of H1, for which the post-probability of H0, given by Equation (3), is equal to the p-value, given by Equation (2). Relaxing the condition that $P(H1\,|-) = P(H0\,|-)$ and noting Equations (1) and (2), we see from Equation (3) that the condition becomes

$$e^{-S} = \frac{e^{-S}}{e^{-S}P(H0\,|-) + P_S(S\,|\,H1)P(H1\,|-)} P(H0\,|-), \tag{10}$$

which leads to the following requirement for $P_S(S\,|\,H1)$:

$$P_S(S\,|\,H1) = \frac{P(H0\,|-)}{P(H1\,|-)}(1 - e^{-S}). \tag{11}$$

It is a requirement of probability theory that the integral of a probability distribution function over the range $0$ *to* $\infty$ should be unity, but the integral of the function in Equation (11) is infinite. We see that **there is no acceptable form of H1 for which the post-probability of H0 is equal to the p-value.** It follows that **the p-value should never be interpreted as the post-probability of the null hypothesis.**

## 4. BASIC REQUIREMENTS CONCERNING H1

Since, as we have seen in the preceding section, the p-value does not provide an acceptable significance estimate for a peak in a power spectrum, we are faced with the challenge of finding a better procedure. We first address this issue by asking "What seems to be the most reasonable and least restrictive form that one can adopt for $P_S(S|H1)$?" Any proposal that is free from the defect shown in Section 3 will be an advance over the use of p-values.

We hope that other analysts will have different and more cogent suggestions, but our current view is that it seems reasonable and minimally restrictive to look for a functional form for $P_S(S|H1)$ that has the following properties:

a . It is nonzero for all values of S,
b . It is a monotonically decreasing function of S,
c . The rate of decrease is as slow as possible, and
d . Its integral is unity.

In seeking the simplest function that meets these needs requirements, (a) and (b) suggest that we adopt an inverse power law for $P_S$:

$$P_S(S|H1) = \frac{A}{(B+S)^\beta}, \qquad (12)$$

where $B > 0$. To meet requirement ( c ), we would like $\beta$ to be as small as possible, but we can meet requirement (d) only if $\beta > 1$.

However, if we adopt $\beta = 1$, the integral diverges but only logarithmically. This suggests that we set an upper limit $S_M$ on the range of powers that we need to consider. Since we rarely encounter power spectra with $S > 20$ (and since if $S > 20$ it is pretty clear that a peak is indeed significant), we propose adopting $\beta = 1$ and $S_M = 20$. Then A is determined by requirement (d), i.e.

$$A = \frac{1}{\ln(1 + S_M / B)} \ . \tag{13}$$

The resulting log-odds of H0 is shown, as a function of S and for the choice $B = 1$, in Figure 1. We also show in this figure the values of the log-odds that we would find by adopting $S_M = 30$. We see that the difference between adopting $S_M = 20$ and $S_M = 30$ (the difference is only 0.05) is negligible. Hence in practice the adoption of a definite upper limit such as $S_M = 20$ is not restrictive and may be ignored.

## 5. RECONCILING BAYESIAN AND FISHERIAN PROCEDURES FOR COMBINING STATISTICS

We now suppose that we have repeated an experiment or an observation, and so obtained two independent measurements of the power, $S_A$ and $S_B$, at a given frequency. From these two values, we may form the corresponding values of the odds, $\Omega_A$ and $\Omega_B$. Then the odds for the two power values, taken together, is given by

$$\Omega_{Y,AB} = \Omega_A * \Omega_B \ , \tag{14}$$

where the subscript Y indicates that this estimate is based on a Bayesian analysis. Hence, introducing the symbol $\Lambda_{Y,AB}$ for the Bayesian estimate of the log-odds on H0 formed by combining $S_A$ and $S_B$, we see that

$$\Lambda_{Y,AB} = \log_{10}\left(\frac{P(S_A | H0)}{P(S_A | H1)}\right) + \log_{10}\left(\frac{P(S_B | H0)}{P(S_B | H1)}\right). \tag{15}$$

Whether or not we adopt the p-value for power-spectrum significance estimation (which in our view we should not do), it is a familiar statistic that most statisticians will continue to look at. It is therefore interesting to reconsider the problem of combining the results of two power-spectrum measurements, as expressed in terms of p-values. Fisher (1938) pointed out that one may derive a procedure for combining p-values from the fact that the sum of a number of chi-square values itself satisfies the chi-square distribution with the appropriate number of degrees of freedom. Noting that, for a single measurement

$$p-value = \exp\left(-\tfrac{1}{2}\chi^2\right), \tag{16}$$

Rosenthal (1984) has expressed Fisher's rule as follows: *If one has a set of independent p-values, the sum of their natural logarithms, multiplied by -2, is a chi-square statistic with twice as many degrees of freedom as there are p-values.* On using Equation (16), and on writing

$$Z = S_1 + ... + S_N, \tag{17}$$

Rosenthal's prescription may be written as

$$PV(Z) = chi2tail(2Z, 2N). \tag{18}$$

in which the terms "chi2tail" denotes the tail of the chi-square function, where "tail(x)" means "1 - cdf(x)." If we write

$$PV(Z) = \exp(-G), \tag{19}$$

we find that Equation (18) leads to

$$G = Z - \ln\left(1 + Z + \tfrac{1}{2}Z^2 + \ldots + \tfrac{1}{(N-1)!}Z^{(N-1)}\right). \tag{20}$$

This expression is in fact identical to the formula for the "Combined Power Statistic" (Sturrock, Scargle, Walther and Wheatland, 2005) previously derived by quite different arguments.

We again consider the process of combining just two power estimates, $S_A$ and $S_B$. If we now write

$$Z_{AB} = S_A + S_B, \tag{21}$$

the statistic shown in Equation (20) becomes

$$G_{AB} = Z_{AB} - \ln(1 + Z_{AB}). \tag{22}$$

We may now form from $G_{AB}$ another estimate of the log-odds derived from the two power values,

$$\Lambda_{F,AB} = \log_{10}\left(\frac{P_S(S_A + S_B - \ln(1 + S_A + S_B)|H0)}{P_S(S_A + S_B - \ln(1 + S_A + S_B)|H1)}\right), \tag{23}$$

where the subscript F indicates that this estimate is based on a Fisherian analysis.

The ideal solution would be to find a functional form for $P_S(S|H1)$ which guarantees that $\Lambda_{F,AB} = \Lambda_{Y,AB}$ for all values of $S_A$ and $S_B$. We may adopt, as a convenient general representation, the ratio of two polynomials:

$$P_S(S|H1) = \frac{a_0 + a_1 S + a_2 S^2 + \ldots}{b_0 + b_1 S + b_2 S^2 + \ldots}. \tag{24}$$

It is necessary that the function have a finite integral (to be set equal to unity), and that the function tend to zero as S tends to infinity.

It is in principle possible to search for a wide range of values $a_0, a_1, a_2, ..., b_0, b_1, b_2, ...$, to find the combination that leads to the best match between $\Lambda_{F,AB}$ and $\Lambda_{Y,AB}$. For instance, we may search for values that minimize the root-mean-square difference between the log-odds as computed (for many pairs of values of $S_A$ and $S_B$) by the Bayesian and the Fisherian procedures,

$$\Delta = \left[ \frac{1}{N_A N_B} \sum_{S_A, S_B} (\Lambda_{Y,AB} - \Lambda_{F,AB})^2 \right]^{1/2}, \tag{25}$$

where $N_A$, $N_B$ are the number of values of $S_A$ and $S_B$ involved in the calculation.

The simplest form of Equation (24) that can meet the basic requirements following that equation is

$$P_S(S|H1) = \frac{a_0}{b_0 + b_1 S + b_2 S^2}, \tag{26}$$

where $b_2$ is nonzero. However, we may also consider the even simpler form

$$P_S(S|H1) = \frac{a_0}{b_0 + b_1 S}, \tag{27}$$

since – as we noted in Section 5 - the integral of this function diverges only logarithmically. We therefore again assign an upper value $S_M$ (such as 20) to the range of powers we need to consider, and require only that

$$\int_0^{S_M} \frac{a_0 dS}{b_0 + b_1 S} = 1. \tag{28}$$

Hence we find that $P_S$ may be expressed as

$$P_S(S|H1) = \frac{1}{\ln(1+S_M/b)} \frac{1}{b+S}, \tag{29}$$

which is essentially identical to the form proposed in Section 5.

For the choice $S_A, S_B = 2, 4, ..., 20$, we find that $\Delta$ has the minimum value 0.094 for $b = 0.52$. For this choice, Equation (29) becomes

$$P_S(S|H1) = \frac{0.41}{1.92 + S}. \tag{30}$$

We see from Equations (15), (23) and (30) that, for our approximate analysis,

$$\Lambda_{Y,AB} = \log_{10}\left[2.44(1.92+S_A)e^{-S_A}\right] + \log_{10}\left[2.44(1.92+S_B)e^{-S_B}\right], \tag{31}$$

and

$$\Lambda_{F,AB} = \log_{10}\left[2.44\{1.92 + S_A + S_B - \ln(1+S_A+S_B)\}(1+S_A+S_B)\exp(-S_A - S_B)\right]. \tag{32}$$

We show in Figure 2 the comparison of $\Lambda_F$ and $\Lambda_Y$ for $S_A$ in the range 0 to 20, and for four values of $S_B$: 2, 4, 8 and 16. We see that, for most combinations of powers, the agreement is quite good. For $S_1, S_2 \geq 4$, the maximum discrepancy is only about 0.25.

## 6. EXAMPLES AND DISCUSSION

Pending a more securely based analysis, we may use the following expression for the odds on H0:

$$\Omega(H0|S) = 2.44(1.92+S)e^{-S}. \tag{33}$$

We show the log-odds of H0 as a function of S and (for comparison) the logarithm of $e^{-S}$ in Figure 3. We also show a list of the comparative values in Table 1. We see that (as expected) significance estimates based on our Bayesian analysis are substantially more conservative than the more familiar p-value estimates.

As an example, we consider the application of the Bayesian approach to the power spectrum derived from Super-Kamiokande (Fukuda et al., 2001, 2002; Fukuda, 2003) solar neutrino data (Sturrock, Caldwell, Scargle, and Wheatland, 2005). We found that there was a notable peak (at frequency $9.43\, yr^{-1}$) with power $S = 11.67$. The corresponding p-value is $8.5\, 10^{-6}$ but the odds value given by Equation (33) is
$2.8\, 10^{-4}$, larger by a factor of 33.

However, we usually need to evaluate the significance of the largest peak in an array of peaks. Scargle (1982) has shown that, on the basis of the null hypothesis H0, the p-value significance estimate of the biggest peak $S_M$ out of M peaks is given by

$$pv = \left[1 - \left(1 - e^{-S_M}\right)\right]^M. \tag{34}$$

Since Equation (33) may be expressed in terms of the p-values as follows,

$$\Omega(H0 \mid pv) = 2.44\left[1.92 - \ln(pv)\right]pv. \tag{35}$$

we find that an odds-measure of the significance of the biggest peak out of M peaks is given by

$$\Omega(H0 \mid S_M, M) = 2.44\left[1.92 - M\ln\left(1 - \left(1 - e^{-S_M}\right)\right)\right]\left[1 - \left(1 - e^{-S_M}\right)\right]^M. \tag{36}$$

For the Kamiokande power spectrum, we find that, for a search band $1 - 36\, yr^{-1}$ (the widest band compatible with no aliasing) $M = 126$. We find from Equation (34) that the "false-alarm" p-value is 0.0011. However, we find from Equation (36) that this corresponds to an odds-value on H0 of 0.023, a much more conservative estimate.

As a second example, we consider the recent combined analysis (Sturrock 2008b) of Homestake (Davis, 1996; Davis, Harmer, and Hoffman, 1968; Cleveland et al., 1998) and GALLEX (Anselmann *et al*., 1993, 1995; Hampel *et al.,* 1996, 1999) radiochemical data and ACRIM irradiance data (Willson 1979, 2001; www.acrim.com)). We have formed the joint power statistic (JPS; Sturrock, Scargle, Walther, and Wheatland, 2005) from four independent datasets: the Homestake neutrino data, the GALLEX neutrino data, ACRIM data for the Homestake time interval, and ACRIM data for the GALLEX time interval. For four power spectra, the joint power statistic is given - to sufficient accuracy - by

$$J = \frac{3.88\, Y^2}{1.27 + Y}, \qquad (37)$$

where

$$Y = \left(S_1 * S_2 * S_3 * S_4\right)^{1/4}. \qquad (38)$$

We find a striking peak (J = 40.87) in the JPS spectrum at 11.85 year$^{-1}$. We interpret this frequency as the synodic rotation frequency of the core, corresponding to a sidereal rotation frequency of 12.85 year$^{-1}$, or 407 nHz. Since the JPS is designed to have the same exponential distribution as the individual powers from which it is formed, we may evaluate its significance by the conventional procedure and by our Bayesian procedure.

Adopting a search band of 10 – 15 year$^{-1}$ for rotational frequencies, we find that there are 66 peaks in the JPS in that band. The false-alarm formula of Equation (34) then leads to the estimate of 1.2 10$^{-16}$ for obtaining the value J = 40.87 or higher by chance. On the other hand, Equation (36) leads to an odds value of 6.6 10$^{-13}$ for hypothesis H0. (The probability has essentially the same value.) This is more conservative than the false-alarm frequency estimate, but it still represents very strong evidence that the neutrino and irradiance data are subject to a common periodic signal.

## Acknowledgements

We wish to thank Alexander Kosovichev and Guenther Walther for helpful discussions related to this work, which was supported by NSF Grant AST-0607572.

Table 1. Comparison of the Log-Odds of H0, derived from Equation (33), and the corresponding values of $Log10(e^{-S})$

| S | Log Odds | $Log10(e^{-S})$ |
|---|---|---|
| 0 | 0.67 | 0.00 |
| 1 | 0.42 | -0.43 |
| 2 | 0.11 | -0.87 |
| 3 | -0.22 | -1.30 |
| 4 | -0.58 | -1.74 |
| 5 | -0.94 | -2.17 |
| 6 | -1.32 | -2.61 |
| 7 | -1.70 | -3.04 |
| 8 | -2.09 | -3.47 |
| 9 | -2.48 | -3.91 |
| 10 | -2.88 | -4.34 |
| 11 | -3.28 | -4.78 |
| 12 | -3.68 | -5.21 |
| 13 | -4.08 | -5.65 |
| 14 | -4.49 | -6.08 |
| 15 | -4.90 | -6.51 |
| 16 | -5.31 | -6.95 |
| 17 | -5.72 | -7.38 |
| 18 | -6.13 | -7.82 |
| 20 | -6.54 | -8.25 |

FIGURES

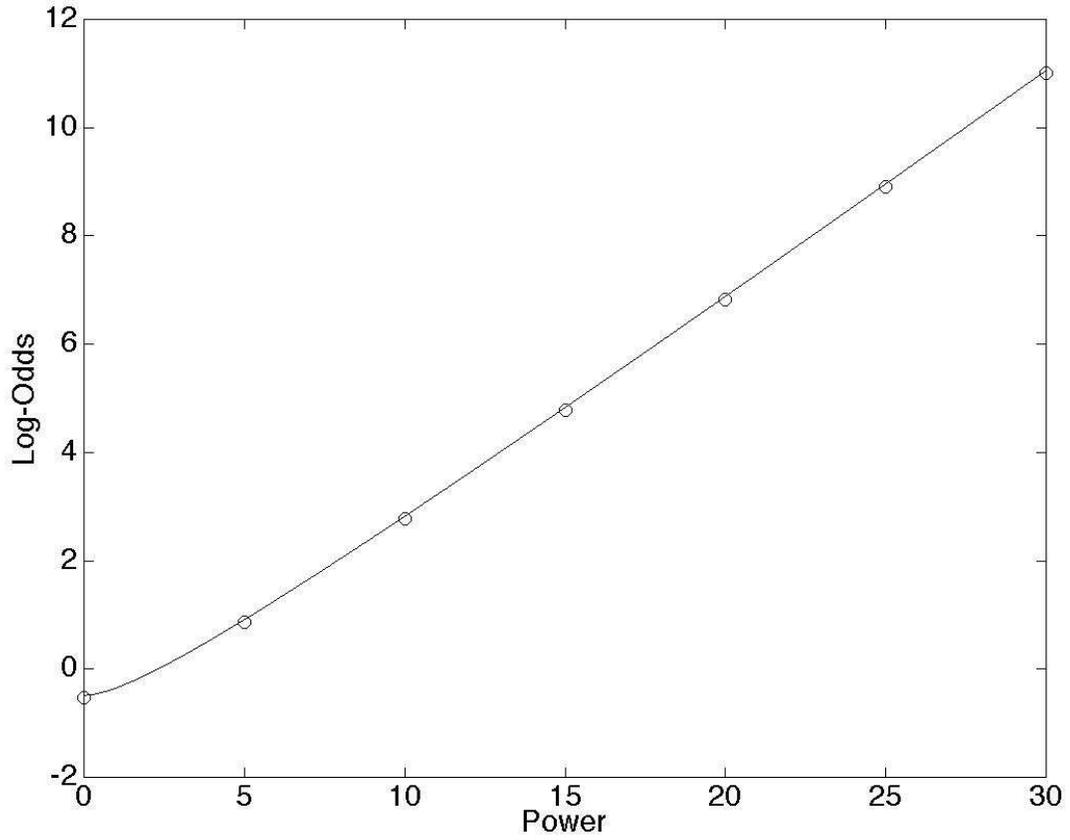

Figure 1. The log-odds of H1 as a function of S for the choice $\beta = 1$, $B = 1$ and $S_M = 20$. The figure also shows, as circles, the values of the log-odds computed with $S_M = 30$.

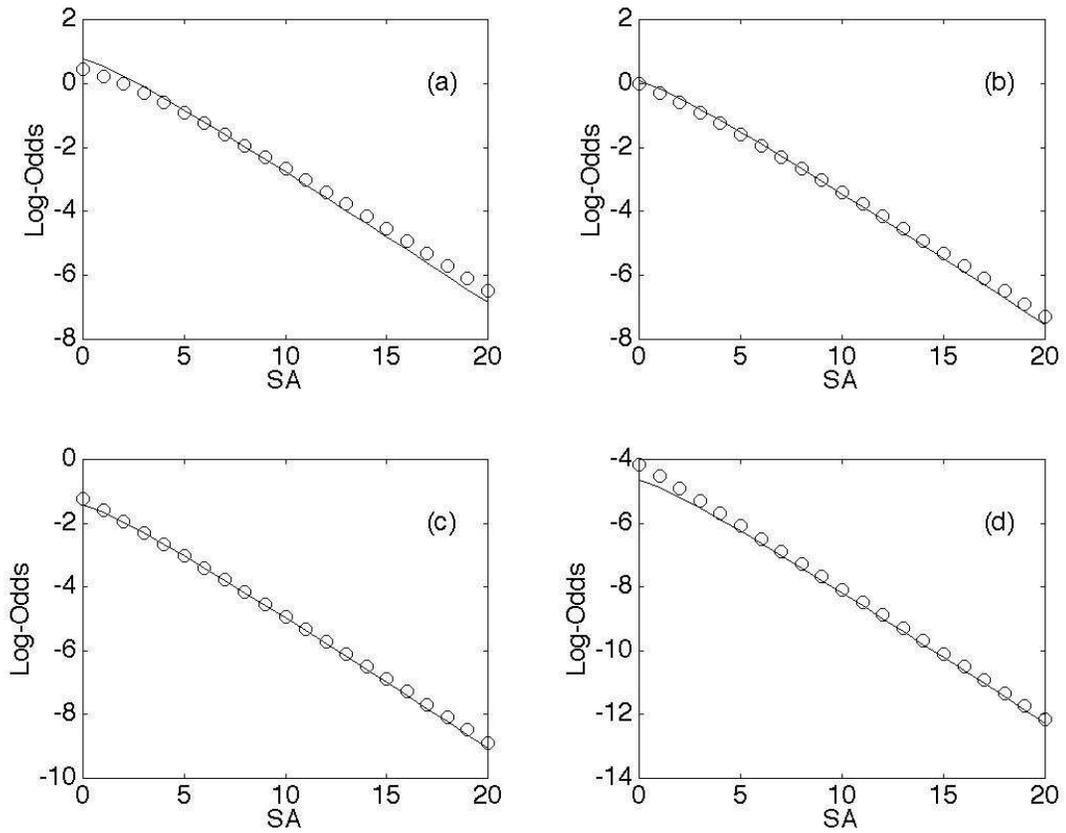

Figure 2. The Bayesian estimate of the log-odds for H0 for values of SA shown in he abscissa, and for four values of SB: (a) SB = 2, (b) SB = 4, (c) SB = 8, (d) SB = 16. Samples of the corresponding Fisher estimates (Equation (29)) are shown as circles.

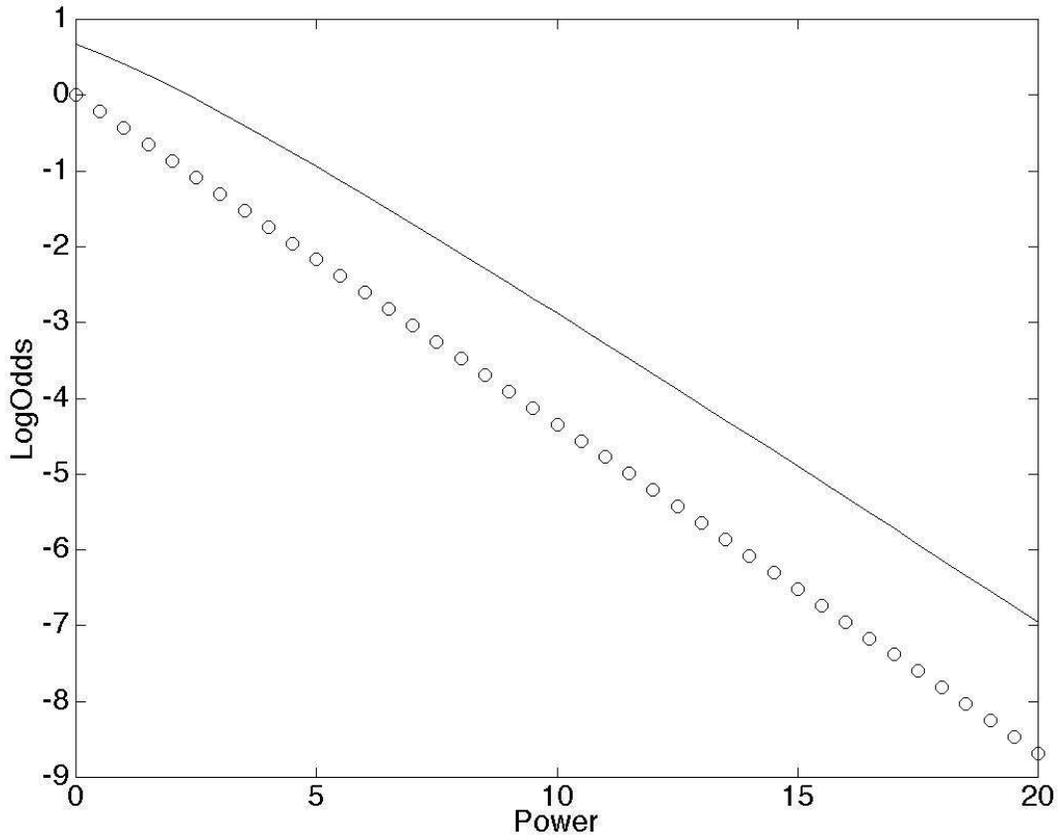

Figure 3. The log-odds of H0 as a function of S as given by Equation (33). The figure also shows, as circles, the logarithm of $\exp(-S)$.